\documentclass[12pt,a4paper]{article}

\title{}
\author{}
\date{}

\usepackage{amsmath}
\usepackage{amsfonts}
\usepackage{amssymb}
\usepackage{latexsym}

\makeatletter
\@addtoreset{equation}{section}
\makeatother

\newcommand{\be}{\begin{equation}}
\newcommand{\ee}{\end{equation}}
\newcommand{\bea}{\setlength\arraycolsep{2pt} \begin{eqnarray}}
\newcommand{\eea}{\end{eqnarray}}
\newcommand{\nnr}{\nonumber \\}

\newcommand{\eq}[1]{(\ref{#1})}

\newcommand{\fr}{\frac}
\newcommand{\tf}{\tfrac}

\newcommand{\df}{\textrm{d}}
\newcommand{\expe}[1]{\textrm{e}^{#1}}
\newcommand{\pd}{\partial}
\newcommand{\sr}{\sqrt}

\newcommand{\gvf}{\varphi}

\newcommand{\im}{\textrm{i}}

\newcommand{\bbR}{\mathbb{R}}

\newcommand{\wh}{\widehat}

\begin{document}

\thispagestyle{empty}

\begin{flushright}
CCTP-2016-13\\
CCQCN-2016-162
\end{flushright}
\vspace*{100pt}
\begin{center}
\textbf{\Large{Hidden symmetries of black holes in five-dimensional supergravity}}\\
\vspace{50pt}
\large{David D. K. Chow}
\end{center}

\begin{center}
\textit{Crete Center for Theoretical Physics and\\
Crete Center for Quantum Complexity and Nanotechnology,\\
Department of Physics, University of Crete, 71003 Heraklion, Greece}\\
{\tt dchow@physics.uoc.gr}\\
\vspace{30pt}
{\bf Abstract\\}
\end{center}
We consider a general charged, rotating black hole in five-dimensional $STU$ supergravity, and show that its six-dimensional Kaluza--Klein lift admits a Killing--Yano 3-form with torsion.  This underlies its known Killing tensors in five dimensions, and is related to the separability of torsion-modified Dirac equations.  In the generalization to gauged supergravity, we present a five-dimensional Killing--Yano 3-form with torsion when two of the gauge fields are equal, and a Killing--St\"{a}ckel tensor in the general 3-charge Wu solution.
\newpage


\section{Introduction}


The 3-charge Cveti\v{c}--Youm solution \cite{Cvetic:1996xz} is a general family of charged, rotating black hole solutions of 5-dimensional $STU$ supergravity.  Through U-duality, it generates general black holes of maximal 5-dimensional supergravity \cite{Cvetic:1996zq}.  In Einstein frame, the massless Klein--Gordon equation separates \cite{Cvetic:1997uw, Cvetic:1998xh}, and similarly the Hamilton--Jacobi equation for null geodesics separates \cite{Davis:2006cs}.  However, the geometry admits only three commuting Killing vectors, corresponding to time translation and rotations in two independent 2-planes.  For a 5-dimensional spacetime, this alone is not enough symmetry to separate variables, so there is a hidden symmetry.

Separability has been used in studying physical properties of these black holes and related solutions.  Greybody factors of the black holes were studied in \cite{Cvetic:1997uw, Cvetic:1998xh}.  For supersymmetric microstate geometries that can be obtained as a limit, separability of the Klein--Gordon equation in 6 dimensions was used in identifying dual conformal field theory (CFT) states \cite{Giusto:2004id, Giusto:2004ip}.  Recently, separability has been used in studying quasinormal modes in these supersymmetric microstate geometries, from which it has been argued that they are unstable \cite{Eperon:2016cdd}.  Non-supersymmetric microstate geometries can also be obtained as a limit \cite{Jejjala:2005yu}, and again separability helps to show instability \cite{Cardoso:2005gj}, and study dual CFT states \cite{Chakrabarty:2015foa}.  Subtracted geometries \cite{Cvetic:2011hp}, which differ from the black hole geometries through different asymptotic behavior, but have the same horizon behavior, can also be obtained as a limit \cite{Cvetic:2012tr}.  By construction, they preserve separability of the Klein--Gordon equation.

Separability is also used in the construction of exact solutions.  Asymptotically anti-de Sitter (AdS) black holes in gauged supergravities are of interest because of the AdS/CFT correspondence, which is best understood in 5 dimensions.  Separability has been the main guide in finding generalizations of the 3-charge Cveti\v{c}--Youm solution in gauged supergravity, culminating in the most general 3-charge Wu solution \cite{Wu:2011gq}.

Geometrically, separability is related to Killing tensors, which can be symmetric or antisymmetric.  Symmetric rank-2 Killing--St\"{a}ckel (KS) tensors $K_{a b} = K_{(a b)}$ are related to the separability of the Hamilton--Jacobi equation for geodesic motion and the Klein--Gordon equation.  The 4-dimensional Kerr solution not only admits a rank-2 KS tensor \cite{Walker:1970un}, but also an antisymmetric Killing--Yano (KY) 2-form $Y_{a b} = Y_{[a b]}$  \cite{Floyd, Penrose:1973um}.  The KY 2-form is a ``square root'' of the rank-2 KS tensor, $K_{a b} = Y{^c}{_a} Y_{c b}$, related to separability of the Dirac equation, and so is a more fundamental geometrical object.  The study of analogous Killing tensors for higher-dimensional Kerr--NUT--AdS solutions of Einstein gravity was initiated in \cite{Frolov:2002xf, Frolov:2003en} for the 5-dimensional Myers--Perry solution.  The general situation in arbitrary dimensions is now well established; for example, see \cite{Kubiznak:2008qp, Yasui:2011pr}.  Killing tensors correspond to hidden symmetries, i.e.\ symmetries of phase space, which can be found more generally across a wide range of physical systems; for example, see \cite{Cariglia:2014ysa}.

For the known charged generalizations in supergravity, there are two main differences, associated with the two fields that generically occur in string theory: the scalar dilaton and the Kalb--Ramond 3-form field strength.  The dilaton relates two significant conformal frames: string frame and Einstein frame.  For the known examples, the string frame admits KS tensors, whereas the Einstein frame generically only admits their conformal generalizations.  In some examples, there are analogues of Killing--Yano $p$-forms, but they typically involve a connection that has a torsion identified with the 3-form field.

There have been a number of previous studies of Killing tensors for the 3-charge Cveti\v{c}--Youm solution and its generalizations to gauged supergravity.  Conformal KS tensors in Einstein frame were given in \cite{Davis:2006cs}, and KS tensors in string frame were given in \cite{Chow:2008fe}.  For special cases, including gauging, KYT 3-forms in string frame were studied in \cite{Wu:2009cn, Kubiznak:2009qi, Wu:2009ug, Houri:2010fr}.  

Since we are considering black holes in string theory, it is natural to examine their properties in different dimensions.  The construction of the solutions often uses coordinate transformations with the extra dimensions.  We shall see that the underlying symmetries of a solution can become clearer from a higher-dimensional perspective.	Physically, there may be singularity resolution in higher dimensions, and the tools of string theory can be used to understand black hole microscopics.  This has motivated recent work on studying Killing tensors for Kaluza--Klein lifts of black holes \cite{Houri:2012su, Chervonyi:2015ima, Chow:2015sfk}.

In this paper, we show that the general 3-charge Cveti\v{c}--Youm solution, when lifted to 6 dimensions, admits a string frame KYT 3-form.  In a suitable orthonormal frame, essentially given previously in \cite{Mei:2007bn}, the KYT 3-form has a simple expression.  This induces a rank-2 KS tensor in 6 dimensions, which projects onto a rank-2 KS tensor in 5 dimensions \cite{Carter:1989bs}.  The KYT 3-form therefore provides a deeper origin of the known properties of the 3-charge Cveti\v{c}--Youm solution.  Generalizing to gauged supergravity, we present a 5-dimensional KYT 3-form in the special case of the Mei--Pope solution \cite{Mei:2007bn}, which has two gauge fields equal, and a rank-2 KS tensor in the general 3-charge Wu solution \cite{Wu:2011gq}.

KYT forms are related to symmetry operators for Dirac operators that are modified by torsion, leading to separability of the corresponding modified Dirac equation.  However, when there is torsion, the construction of the symmetry operator requires not only the KYT form, but also the vanishing of certain anomaly terms \cite{Houri:2010qc}.  In particular, there is a ``classical'' anomaly, so-called because it can be seen in a pseudo-classical theory of a spinning particle \cite{Kubiznak:2009qi}, and a ``quantum'' anomaly, which only appears at the operator level.  We show that the two anomaly terms are related by Hodge duality.  For the solutions we consider that admit KYT 3-forms, both anomalies can be cancelled, and so symmetry operators for the modified Dirac operator can be constructed.

The outline of this paper is as follows.  In Section 2, we consider asymptotically flat black holes in ungauged $STU$ supergravity, presenting Killing tensors in 5 and 6 dimensions.  In Section 3, we consider asymptotically AdS black holes in gauged $STU$ supergravity, presenting 5-dimensional Killing tensors.  Section 4 studies symmetry operators for torsion-modified Dirac operators, which we show can be constructed for the solutions admitting KYT 3-forms.  We conclude in Section 5.  An Appendix provides the explicit formulae to see variable separation for the general 3-charge Wu solution.


\section{Ungauged supergravity}


%
%

5-dimensional $STU$ supergravity can be considered as minimal $\mathcal{N} = 2$ supergravity coupled to two vector multiplets.  The bosonic fields are the metric $g_{a b}$, three $\textrm{U}(1)$ gauge fields $A_I$, $I = 1, 2, 3$, and two dilatons $\gvf_i$, $i = 1, 2$.  The Einstein frame Lagrangian is
\be
\mathcal L = R \star 1 - \fr{1}{2} \sum_{i = 1}^2 \star \df \gvf_i \wedge \df \gvf_i - \fr{1}{2} \sum_{I = 1}^3 X_I^{-2} \star F_I \wedge F_I + F_1 \wedge F_2 \wedge A_3 ,
\label{Lagrangian}
\ee
where $F_I = \df A_I$, and the scalar combinations $X_I$ are
\begin{align}
X_1 & = \expe{- \gvf_1 / \sr{6} - \gvf_2 / \sr{2}} , & X_2 & = \expe{- \gvf_1 / \sr{6} + \gvf_2 / \sr{2}} , & X_3 & = \expe{2 \gvf_1/\sr{6}} .
\end{align}
The string frame metric $g_{\textrm{s}}$ and Einstein frame metric $g_{\textrm{E}}$ are related by $g_{\textrm{s}} = X_1 X_2 g_{\textrm{E}}$.
	

\subsection{5-dimensional solution}


To set the conventions, we first present the 6-parameter 3-charge Cveti\v{c}--Youm solution in a standard form in 5 dimensions.  The solution has a mass parameter $m$, two rotation parameters $a$ and $b$, and three electric charge parameters $\delta_I$.  The 5-dimensional string frame metric is concisely expressed in the exquisite form \cite{Wu:2011gq}
\be
\df s^2 = H_3 \bigg( - \df t^2 + (r^2 + a^2) \sin^2 \theta \, \df \phi_1^2 + (r^2 + b^2) \cos^2 \theta \, \df \phi_2^2 + \fr{\rho^2}{\Delta_r} \, \df r^2 + \rho^2 \, \df \theta^2 + \sum_{I = 1}^3 \fr{(1 - 1/H_I) K_I^2}{\prod_{J \neq I} (s_I^2 - s_J^2)} \bigg) ,
\ee
where
\begin{align}
\Delta_r & = \fr{(r^2 + a^2) (r^2 + b^2)}{r^2} - 2 m , & \rho^2 & = r^2 + a^2 \cos^2 \theta + b^2 \sin^2 \theta , & H_I & = 1 + \fr{2 m s_I^2}{\rho^2} ,
\end{align}
and we use the notation $\sinh \delta_I$ and $c_I = \cosh \delta_I$.  The 1-forms $K_I$ are given by
\be
K_1 = s_1 c_1 \, \df t - (a s_1 c_2 c_3 - b c_1 s_2 s_3) \sin^2 \theta \, \df \phi_1 - (b s_1 c_2 c_3 - a c_1 s_2 s_3) \cos^2 \theta \, \df \phi_2 ,
\ee
and analogous expressions for $K_2$ and $K_3$ by permuting the charge parameters $\delta_I$.  The matter fields are given by
\begin{align}
A_I & = \fr{2 m}{H_I \rho^2} K_I , & X_I & = \fr{(H_1 H_2 H_3)^{1/3}}{H_I} .
\end{align}
We can dualize the gauge field strength $F_3$ in favour of a 3-form field strength $H = \df B - \tf{1}{2} (A_1 \wedge F_2 + A_2 \wedge F_1)$ through $H = X_3^{-2} \star F_3$ in Einstein frame.  Our convention is $\varepsilon_{t r \theta \phi_1 \phi_2} = 1$, and we can take
\begin{align}
B & = \fr{m}{\rho^2} \bigg( \fr{1}{H_1} + \fr{1}{H_2} \bigg) [(a s_1 s_2 c_3 - b c_1 c_2 s_3) \sin^2 \theta \, \df t \wedge \df \phi_1 + (b s_1 s_2 c_3 - a c_1 c_2 s_3) \cos^2 \theta \, \df t \wedge \df \phi_2 \nnr
& \quad + (a^2 - b^2) s_3 c_3 \sin^2 \theta \, \cos^2 \theta \, \df \phi_1 \wedge \df \phi_2] + 2 m s_3 c_3 \cos^2 \theta \, \df \phi_1 \wedge \df \phi_2 .
\end{align}


\subsection{6-dimensional solution}


The 6-dimensional string frame metric $\df s_6^2$, 3-form field strength $H_6$, and canonically normalized dilaton $\gvf_6$ are given by the Kaluza--Klein ansatz
\begin{align}
\df s_6^2 & = \df s^2 + \fr{X_1}{X_2} (\df z + A_2)^2 , & H_6 & = H + F_1 \wedge (\df z + A_2) , & \gvf_6 & = \tf{1}{2} (\sr{3} \gvf_1 + \gvf_2) ,
\label{ansatz}
\end{align}
where the 5-dimensional metric $\df s^2$ is in string frame.  The 6-dimensional string frame Lagrangian is
\be
\mathcal{L} = \expe{\sr{2} \gvf_6} (R \star 1 + 2 \star \df \gvf_6 \wedge \df \gvf_6 - \tf{1}{2} \star H_6 \wedge H_6) .
\ee
The 6-dimensional string frame metric $g_{\textrm{s}}$ and Einstein frame metric $g_{\textrm{E}}$ are related by $g_{\textrm{s}} = \expe{- \gvf_6/\sr{2}} g_{\textrm{E}}$, and so $g_{\textrm{s}} = \sr{H_3/H_1} g_{\textrm{E}}$.

Define the new parameters
\begin{align}
\wh{a} & = a c_3 - b s_3 , & \wh{b} & = b c_3 - a s_3 ,
\end{align}
the shifted coordinates
\begin{align}
\wh{r}^2 & = r^2 + 2 m s_3^2 + 2 a b s_3 c_3 - (a^2 + b^2) s_3^2 , \nnr
\wh{y}^2 & = a^2 \cos^2 \theta + b^2 \sin^2 \theta - 2 a b s_3 c_3 + (a^2 + b^2) s_3^2 ,
\end{align}
and the functions
\begin{align}
\wh{R} & = (\wh{r}^2 + \wh{a}^2) (\wh{r}^2 + \wh{b}^2) - 2 m (1 + 2 s_3^2) \wh{r}^2 + 4 m \wh{a} \wh{b} s_3 c_3 + 4 m^2 s_3^2 c_3^2 , \nnr
\wh{Y} & = (\wh{a}^2 - \wh{y}^2) (\wh{y}^2 - \wh{b}^2) .
\end{align}
Define also the transformed Killing coordinates
\begin{align}
\psi_0 & = C t - \fr{\wh{a}^3}{(\wh{a}^2 - \wh{b}^2)} \phi_1 - \fr{\wh{b}^3}{(\wh{b}^2 - \wh{a}^2)} \phi_2 - S z , & \psi_1 & = \fr{\wh{a}}{\wh{a}^2 - \wh{b}^2} \phi_1 + \fr{\wh{b}}{\wh{b}^2 - \wh{a}^2} \phi_2 , \nnr
\psi_2 & = - \fr{1}{\wh{a} (\wh{a}^2 - \wh{b}^2)} \phi_1 - \fr{1}{\wh{b} (\wh{b}^2 - \wh{a}^2)} \phi_2 , & \psi_3 & =C z - S t ,
\end{align}
where $C = \cosh (\delta_1 - \delta_2)$ and $S = \sinh (\delta_1 - \delta_2)$.  The string frame metric can be written in terms of an orthonormal frame as $\df s^2 = - (e^0)^2 + \sum_{\mu = 1}^5 (e^\mu)^2$, where
\begin{align}
e^0 & = \fr{\sr{\wh{r}^2 + \wh{y}^2} \sr{\wh{R}}}{\wh{r} \sr{\wh{r}^2 + \wh{y}^2 + 2 m (s_1^2 - s_3^2)}} (\df \psi_0 + \wh{y}^2 \, \df \psi_1) , & e^1 & = \fr{\wh{r} \sr{\wh{r}^2 + \wh{y}^2}}{\sr{\wh{R}}} \, \df \wh{r} , \nnr
e^2 & = \fr{\sr{\wh{r}^2 + \wh{y}^2} \sr{\wh{Y}}}{\wh{y} \sr{\wh{r}^2 + \wh{y}^2 + 2 m (s_1^2 - s_3^2)}} [\df \psi_0 - \wh{r}^2 \, \df \psi_1 - 2 m (s_1^2 - s_3^2) \, \df \psi_1] , & e^3 & = \fr{\wh{y} \sr{\wh{r}^2 + \wh{y}^2}}{\sr{\wh{Y}}} \, \df \wh{y} , \nnr
e^4 & = \fr{\wh{a} \wh{b}}{\wh{r} \wh{y}} \bigg( \fr{\wh{r}^2 + \wh{y}^2 + 2 m s_3 c_3 \wh{y}^2 / \wh{a} \wh{b}}{\wh{r}^2 + \wh{y}^2 + 2 m (s_1^2 - s_3^2)} (\df \psi_0 + \wh{y}^2 \, \df \psi_1) - \wh{r}^2 (\df \psi_1 + \wh{y}^2 \, \df \psi_2) \bigg) , \nnr
e^5 & = \df \psi_3 + \fr{2 m s_1 c_1}{\wh{r}^2 + \wh{y}^2 + 2 m (s_1^2 - s_3^2)} (\df \psi_0 + \wh{y}^2 \, \df \psi_1) .
\label{vielbeins}
\end{align}
The metric inverse can be written in terms of the corresponding dual vector fields as $(\pd / \pd s)^2 = - (e_0)^2 + \sum_{\mu = 1}^5 (e_\mu)^2$, where
\begin{align}
e_0 & = \fr{1}{\wh{r} \sr{\wh{r}^2 + \wh{y}^2} \sr{\wh{R}}} \bigg[ \wh{r}^4 \, \pd_{\psi_0} + \wh{r}^2 \, \pd_{\psi_1} + \pd_{\psi_2} + 2 m \bigg( (s_1^2 - s_3^2) \wh{r}^2 \pd_{\psi_0} + \fr{s_3 c_3}{\wh{a} \wh{b}} \, \pd_{\psi_2} - s_1 c_1 \wh{r}^2 \, \pd_{\psi_3} \bigg) \bigg] , \nnr
e_1 & = \fr{\sr{\wh{R}}}{\wh{r} \sr{\wh{r}^2 + \wh{y}^2}} \, \pd_{\wh{r}} , \qquad e_2 = \fr{1}{\wh{y} \sr{\wh{r}^2 + \wh{y}^2} \sr{\wh{Y}}} (\wh{y}^4 \, \pd_{\psi_0} - \wh{y}^2 \, \pd_{\psi_1} + \pd_{\psi_2}) , \nnr
e_3 & = \fr{\sr{\wh{Y}}}{\wh{y} \sr{\wh{r}^2 + \wh{y}^2}} \, \pd_{\wh{y}} , \qquad e_4 = - \fr{1}{\wh{a} \wh{b} \wh{r} \wh{y}} \, \pd_{\psi_2} , \qquad e_5 = \pd_{\psi_3} .
\label{inversevielbeins}
\end{align}
These transformations and orthonormal frame are essentially those of \cite{Mei:2007bn}.  The string frame metric determinant is given by $\sr{-g} = \wh{a} \wh{b} \wh{r} \wh{y} (\wh{r}^2 + \wh{y}^2)^2 / [\wh{r}^2 + \wh{y}^2 + 2 m (s_1^2 - s_3^2)]$.  The components $(\wh{r}^2 + \wh{y}^2) g^{a b}$ are clearly separable as functions of $\wh{r}$ plus functions of $\wh{y}$, which leads to the separability of the Hamilton--Jacobi equation for geodesic motion and the Einstein frame massless Klein--Gordon equation.  The 3-form field strength can be expressed as
\begin{align}
H_6 & = \fr{4 m}{(\wh{r}^2 + \wh{y}^2) [\wh{r}^2 + \wh{y}^2 + 2 m (s_1^2 - s_3^2)]} \bigg[ s_1 c_1 (\wh{r} e^0 \wedge e^1 + \wh{y} e^2 \wedge e^3) \wedge e^5 \nnr
& \quad + s_3 c_3 (\wh{y} e^0 \wedge e^1 - \wh{r} e^2 \wedge e^3) \wedge e^4 - (s_1^2 - s_3^2) \bigg( \fr{\wh{a} \wh{b}}{\wh{y}} e^0 \wedge e^1 + \fr{\wh{a} \wh{b} + 2 m s_3 c_3}{\wh{r}} e^2 \wedge e^3 \bigg) \wedge e^4 \nnr
& \quad + \fr{(s_1^2 - s_3^2)}{\sr{\wh{r}^2 + \wh{y}^2}} e^0 \wedge e^2 \wedge \bigg( \fr{\wh{r} \sr{\wh{Y}}}{\wh{y}} e^1 + \fr{\wh{y} \sr{\wh{R}}}{\wh{r}} e^3 \bigg) \bigg] .
\label{H6}
\end{align}
Note that the charge parameter $\delta_2$ does not appear explicitly in these expressions because, from the 6-dimensional viewpoint, it can be absorbed by a coordinate change corresponding to a Lorentz boost on the Kaluza--Klein coordinate.


\subsection{6-dimensional Killing tensors}


A Killing--Yano $p$-form with torsion (KYT $p$-form) $Y_{a_1 \ldots a_p} = Y_{[a_1 \ldots a_p]}$ satisfies
\be
\nabla{^T}{_a} Y_{b_1 \ldots b_p} = \nabla{^T}{_{[a}} Y_{b_1 \ldots b_p]} ,
\ee
where the connection $\Gamma{^T}{^a}{_{b c}} = \Gamma{^a}{_{b c}} + \tf{1}{2} T{^a}{_{b c}}$ has contributions from both the Levi-Civita connection $\Gamma{^a}{_{b c}}$ and a torsion $T_{a b c} = T_{[a b c]}$.  Taking the torsion to be the 3-form field strength, $T = H_6$, there is a KYT 3-form
\be
Y = (\wh{y} e^0 \wedge e^1 + \wh{r} e^2 \wedge e^3) \wedge e^4 .
\label{KYT}
\ee

A rank-2 Killing--St\"{a}ckel (KS) tensor $K_{a b} = K_{(a b)}$ satisfies
\be
\nabla_{(a} K_{b c)} = 0 ,
\ee
so that $K^{a b} P_a P_b$ is constant for geodesic motion, where $P_a$ is the momentum.  From a KYT $p$-form $Y_{a_1 \ldots a_p}$, a rank-2 KS tensor is given by
\be
K_{a b} = \tfrac{1}{(p - 1)!} Y{^{c_1 \ldots c_{p - 1}}}{_a} Y_{c_1 \ldots c_{p - 1} b} .
\ee
Therefore, the KYT 3-form \eq{KYT} induces the rank-2 KS tensor
\be
K_{a b} \, \df x^a \, \df x^b = \wh{y}^2 (e^0 e^0 - e^1 e^1) + \wh{r}^2 (e^2 e^2 + e^3 e^3) + (\wh{r}^2 - \wh{y}^2) e^4 e^4 .
\label{KS}
\ee
The 6-dimensional orthonormal frame \eq{vielbeins} matches the eigenvectors of the endomorphism $K{_a}{^b}$.  All Schouten--Nijenhuis brackets of the KS tensor and the Killing vectors vanish, or equivalently the associated constants of motion Poisson commute, so geodesic motion is Liouville integrable.

A rank-2 conformal Killing--St\"{a}ckel (CKS) tensor $Q_{a b} = Q_{(a b)}$ satisfies $\nabla_{(a} Q_{b c)} = q_{(a} g_{b c)}$ for some $q_a$, given in $D$ dimensions by $q_a = \tf{1}{D + 2} (\pd_a Q{^b}{_b} + 2 \nabla_b Q{^b}{_a})$.  If we change conformal frame, such as to Einstein frame, then the string frame KS tensor induces a CKS tensor in the new frame with components given by $Q^{a b} = K^{a b}$.

Note that the 10-dimensional string frame metric is simply a direct product of the 6-dimensional string frame metric and $\bbR^4$, while the 3-form field is unchanged when lifted to 10 dimensions.  Therefore, the 6-dimensional KYT 3-form can be lifted unchanged to 10 dimensions.


\subsection{5-dimensional Killing tensors}
\label{5dKill}


In the general case, the 5-dimensional solution admits a rank-2 KS tensor, but the special case of two equal charges also admits a KYT 3-form.


\subsubsection{General charges}


The 6-dimensional KS tensor projects onto a 5-dimensional KS tensor \cite{Carter:1989bs}.  From the unit normalized 1-form $n_a \, \df x^a = \sr{X_1/X_2} (\df z + A_2)$, we define the projector $h{_a}{^b} = \delta^b_a - n_a n^b$.  Denoting the 6-dimensional KS tensor as $\overline{K}_{a b}$, we obtain the 5-dimensional KS tensor $K_{a b} = h{_a}{^c} h{_b}{^d} \overline{K}_{c d}$.  This agrees with the previously found KS tensor \cite{Chow:2008fe}, up to trivial combinations of products of Killing vectors and the metric.


\subsubsection{Two equal charges}


In the special case that $\delta_1 = \delta_2$, $\df z$ appears only in the vielbein $e^5 = \df z + A_1$.  Therefore the orthonormal frame \eq{vielbeins} is adapted to the Kaluza--Klein lift of a 5-dimensional solution \eq{ansatz}.  The 5-dimensional string frame metric is expressed in terms of the orthonormal frame \eq{vielbeins} as $\df s^2 = - (e^0)^2 + \sum_{\mu = 1}^4 (e^\mu)^2$, and the 5-dimensional 3-form field strength is given by \eq{H6} but setting $e^5 = 0$.  The KYT 3-form \eq{KYT} has no terms involving $e^5$, and so trivially projects to give a KYT 3-form in 5 dimensions, with associated torsion given by the 5-dimensional 3-form $H$.  Conversely, the 5-dimensional KYT 3-form can be lifted to 6 dimensions \cite{Chow:2015sfk}.  KYT 3-forms for the 5-dimensional string frame metric were previously studied for the cases with all three charge parameters equal, $\delta_I = \delta$ \cite{Wu:2009cn}, and with two equal charge parameters and the third charge parameter zero, $\delta_1 = \delta_2$, $\delta_3 = 0$ \cite{Houri:2010fr}.  The induced rank-2 KS tensor is given again in \eq{KS}.  Furthermore, the 5-dimensional Hodge dual gives a closed conformal Killing--Yano 2-form with torsion (CCKYT 2-form) \cite{Kubiznak:2009qi, Houri:2010fr}, in this case
\be
k = \star Y = \wh{r} e^0 \wedge e^1 - \wh{y} e^2 \wedge e^3 .
\ee
$k \wedge k$ is a CCKYT 4-form, whose Hodge dual gives a Killing vector, $\star (k \wedge k) = - (2/\wh{a} \wh{b}) (\pd_{\psi_2})^\flat$.  Generically, $\df k \neq 0$, as in the $\delta_3 = 0$ case \cite{Houri:2010fr}.  However, in the special case that we furthermore have $\delta_1 = \delta_3$, i.e.\ all gauge fields are equal, $k$ is closed and there is a 1-form potential $b$ such that $k = \df b$, from which the remaining Killing vectors can be constructed \cite{Kubiznak:2009qi}.


\section{Gauged supergravity}


Gauged $STU$ supergravity is given by the Lagrangian of the ungauged $STU$ supergravity \eq{Lagrangian} plus a scalar potential,
\be
\mathcal{L}_{\textrm{gauged}} = \mathcal{L} + 4 g^2 \sum_{I = 1}^3 X_I^{-1} \star 1 ,
\ee
where $g$ is the gauge-coupling constant.  The theory is an abelian truncation of the maximal $\mathcal{N} = 8$, $\textrm{SO}(6)$-gauged supergravity.  There are asymptotically AdS charged, rotating black hole solutions, most generally given by the 3-charge Wu solution \cite{Wu:2011gq}.  A 5-dimensional solution can be lifted on $S^5$ to type IIB supergravity using the ansatz of \cite{Cvetic:1999xp}.


\subsection{Two equal charges}


The Mei--Pope solution \cite{Mei:2007bn} is a 6-parameter family of charged, rotating black holes.  Two of the gauge fields are equal, $A_1 = A_2$, while the third gauge field $A_3$ is independent.

Generalizing the discussion in Section \ref{5dKill}, the string frame metric of the solution is expressed in our notation in terms of the orthonormal frame \eq{vielbeins} as $\df s^2 = - (e^0)^2 + \sum_{\mu = 1}^4 (e^\mu)^2$, but with the modified functions
\begin{align}
\wh{R} & = (\wh{r}^2 + \wh{a}^2) (\wh{r}^2 + \wh{b}^2) - 2 m (1 + 2 s_3^2) \wh{r}^2 + 4 m \wh{a} \wh{b} s_3 c_3 + 4 m^2 s_3^2 c_3^2 \nnr
& \quad + g^2 \wh{r}^2 [\wh{r}^2 + \wh{a}^2 + 2 m (s_1^2 - s_3^2)] [\wh{r}^2 + \wh{b}^2 + 2 m (s_1^2 - s_3^2)] , \nnr
\wh{Y} & = (\wh{a}^2 - \wh{y}^2) (\wh{y}^2 - \wh{b}^2) (1 - g^2 \wh{y}^2) .
\end{align}
The expressions for the matter fields are the same as for the ungauged $g = 0$ solution.  There is a KYT 3-form given by the same formula \eq{KYT}, and again the expression for the torsion is given by \eq{H6}, setting $e^5 = 0$.  The rank-2 KS tensor and CCKYT forms are induced as in the discussion of the ungauged solution in Section \ref{5dKill}.

The previous analysis of \cite{Chow:2008fe} instead considered a conformal frame that differed by a factor of $H_3$, so was not string frame when $\delta_3 \neq 0$.  The KYT 3-form was previously studied for the special case when all three charge parameters are equal, $\delta_I = \delta$ \cite{Kubiznak:2009qi, Wu:2009ug}, and when two of the charge parameters are equal and the third charge parameter vanishes $\delta_1 = \delta_2$, $\delta_3 = 0$ \cite{Houri:2010fr}.


\subsection{General charges}


More generally, when all three gauge fields are allowed to be independent, the 3-charge Wu solution gives a 7-parameter family of charged, rotating black holes \cite{Wu:2011gq}.  The components of the string frame metric inverse $g^{a b}$ are given in the Appendix, and have also been discussed in \cite{Birkandan:2014vga}.  The components $H_3 \rho^2 g^{a b}$ again are manifestly separable as functions of $r$ plus functions of $\theta$, which implies the separability of the Hamilton--Jacobi equation for geodesics in string frame, but generically only null geodesics in Einstein frame.  Using the notation of \eq{constants} and \eq{Deltatheta}, a rank-2 KS tensor for the string frame metric is
\begin{align}
K^{a b} \, \pd_a \, \pd_b & = - \fr{\Xi_a \Xi_b}{g^2 \Delta_\theta} \, \pd_t \, \pd_t + \fr{\Xi_a}{\sin^2 \theta} \, \pd_{\phi_1} \, \pd_{\phi_1} + \fr{\Xi_b}{\cos^2 \theta} \, \pd_{\phi_2} \, \pd_{\phi_2} + \Delta_\theta \, \pd_\theta \, \pd_\theta \nnr
& \quad - (a^2 \cos^2 \theta + b^2 \sin^2 \theta) \bigg( \fr{\pd}{\pd s} \bigg) ^2 ,
\end{align}
as expected from the canonical form of $D$-dimensional separable metrics with $D - 2$ independent Killing vectors \cite{Benenti}.  Again, all Schouten--Nijenhuis brackets of the KS tensor and Killing vectors vanish, and so geodesic motion is Liouville integrable.

Recall that the Einstein frame metric $g_{\textrm{E}}$ is related to the string frame metric $g_{\textrm{s}}$ by $g_{\textrm{E}} = (H_1 H_2 / H_3^2)^{1/3} g_{\textrm{s}}$.  In 5-dimensional Einstein frame, the metric determinant is given by $\sr{-g} = (H_1 H_2 H_3)^{1/3} \rho^2 r \sin \theta \, \cos \theta / \Xi_a \Xi_b$.  Converting the metric inverse components in the Appendix to Einstein frame, one sees the multiplicative separation of the Klein--Gordon equation for massless scalars, $\pd_a (\sr{-g} g^{a b} \pd_b \Phi) = 0$, by using the ansatz $\Phi = R(r) \Theta(\theta) \expe{\im (- \omega t + \Phi_1 \phi_1 + \Phi_2 \phi_2)}$, which has been analyzed in further detail \cite{Birkandan:2014vga}.


\section{Dirac equation with torsion}


KYT forms and their Hodge dual CCKYT forms are associated with off-shell first-order symmetry operators for the torsion-modified Dirac operator $\mathcal{D} = \gamma^a \nabla_a - \tf{1}{24} T_{a b c} \gamma^{a b c}$.  More generally, conformal Killing--Yano forms with torsion (CKYT forms) are associated with on-shell symmetry operators.  However, these operators rely on further conditions: the vanishing of a ``classical'' and a ``quantum'' anomaly \cite{Houri:2010qc}.

We follow the conventions of \cite{Kubiznak:2010ig}, but simplify all formulae by changing the normalization of the contracted exterior product.  For a $p$-form $\alpha$ and a $q$-form $\beta$, we define
\be
(\alpha \wedge_n \beta)_{c_1 \ldots c_{p + q - 2 n}} = \fr{(p + q - 2 n)!}{n! (p - n)! (q - n)!} \alpha{^{a_1 \ldots a_n}}{_{[c_1 \ldots c_{p - n}}} \beta_{| a_1 \ldots a_n | c_{p - n + 1} \ldots c_{p + q - 2 n}]} .
\ee
With this normalization, the torsion-modified exterior derivative and codifferential for a $p$-form $\omega$ are $\df^T \omega = \df \omega - T \wedge_1 \omega$ and $\delta^T \omega = \delta \omega - T \wedge_2 \omega$.

For a $D$-dimensional CKYT $p$-form $\omega$ the $(p + 2)$-form classical anomaly and $(p - 2)$-form quantum anomaly are \cite{Houri:2010qc}
\begin{align}
A_{(\textrm{c})} (\omega) & = \fr{\df (\df^T \omega)}{p + 1} - \fr{T \wedge \delta^T \omega}{D - p + 1} - \fr{1}{2} \df T \wedge_1 \omega , \nnr
A_{(\textrm{q})} (\omega) & = \fr{\delta (\delta^T \omega)}{D - p + 1} - \fr{T \wedge_3 \df^T \omega}{p + 1} + \fr{1}{2} \df T \wedge_3 \omega .
\end{align}
Using the relation $\alpha \wedge_r \star \beta = (-1)^{p (q + r + 1)} \star (\alpha \wedge_{p - r} \beta)$ for a $p$-form $\alpha$ and a $q$-form $\beta$ \cite{Houri:2010fr}, we see that a CKYT $p$-form $\omega$ has the ``classical/quantum'' duality
\be
A_{(\textrm{c})} (\star \omega) = - \star A_{(\textrm{q})} (\omega) .
\ee
Therefore, if both anomalies vanish for $\omega$, then they both vanish for $\star \omega$.

We can construct a first-order off-shell symmetry operator if the inhomogeneous form $A_{(\textrm{c})} + A_{(\textrm{q})} - \df f + \delta \epsilon$ vanishes, where $f$ is a scalar and $\epsilon$ is a $D$-form \cite{Kubiznak:2010ig}.  Generically, we need $A_{(\textrm{c})} = 0$ and $A_{(\textrm{q})} = 0$, with $f = 0$ and $\epsilon = 0$ also.  In the special case that $p = 3$, the quantum anomaly $A_{(\textrm{q})}$ can be non-zero, but must be exact, $A_{(\textrm{q})} = \df f$.  Similarly, in the Hodge dual case that $p = D - 3$, the classical anomaly $A_{(\textrm{c})}$ can be non-zero, but must be coexact, $A_{(\textrm{c})} = - \delta \epsilon$.  The ``classical/quantum'' duality continues to hold with $\epsilon = (-1)^D \star f$.

Consider the KYT 3-form $Y$ of the 6-dimensional solution of ungauged supergravity with two equal charges.  This satisfies $\delta^{H_6} Y = 0$, and since $\df H_6 = 0$ on-shell, we have simply $A_{(\textrm{c})} (Y) = \tf{1}{4} \df (\df^{H_6} Y)$ and $A_{(\textrm{q})} (Y) = - \tf{1}{4} H_6 \wedge_3 \df^{H_6} Y$, which gives
\begin{align}
A_{(\textrm{c})} (Y) & = 0 , & A_{(\textrm{q})} (Y) & = \df \bigg( \fr{2 m s_3 c_3}{\wh{r}^2 + \wh{y}^2} \bigg) .
\label{anomalyY}
\end{align}
The anomaly conditions are satisfied, so one can construct an off-shell symmetry operator, whose explicit expression is given by the results of \cite{Kubiznak:2010ig}.  Similarly, one can construct a symmetry operator using the Hodge dual CCKYT 3-form.

Simlarly, consider the KYT 3-form $Y$ of the 5-dimensional solution of ungauged or gauged supergravity with two equal charges.  The KYT 3-form $Y$ again satisfies \eq{anomalyY}, so we can construct symmetry operators out of $Y$ and its Hodge dual CCKYT 2-form $k$.  We can further generalize the Dirac operator to $\mathcal{D} = \gamma^a \nabla_a + \im A_a \gamma^a - \tf{1}{24} T_{a b c} \gamma^{a b c}$, where $A = A_1 = A_2$ is the $\textrm{U}(1)$ gauge field.  This gives a further anomaly condition \cite{Kubiznak:2010ig}, $(\df A) \wedge_1 Y = 0$, which is satisfied in this example.


\section{Discussion}


We have seen that the Killing tensors for the general 3-charge Cveti\v{c}--Youm solution can be derived from a 6-dimensional KYT 3-form.  Since the KYT 3-form trivially lifts to 10 dimensions, one may prefer to think of the 10-dimensional KYT 3-form as being the more fundamental Killing tensor.  This viewpoint is reinforced by the fact that Killing tensors single out string frame as the preferred conformal frame.  Similar results for other dimensions will be discussed elsewhere \cite{chow}.

We have also highlighted the significance of the 5-dimensional solution with two equal gauge fields and a third independent gauge field.  This case is particularly relevant to 5-dimensional subtracted geometries \cite{Cvetic:2011hp}, which can be obtained as a scaling limit \cite{Cvetic:2012tr}, and so possess KYT 3-forms that might be related to conformal symmetry.  It may also be possible to find larger families of geometries admitting KYT forms, in five and other dimensions, extending the examples of \cite{Houri:2012eq}.

KYT forms correspond to the separability of torsion-modified Dirac equations.  The explicit construction of symmetry operators and separation could be investigated in further detail for the solutions considered here.  There are proposed correspondences between black holes and CFTs that are based on conformal symmetry in the Klein--Gordon equation, but there has been much less study of higher-spin fields.

It would be interesting to understand if the general 3-charge Wu solution of gauged supergravity admits further Killing tensors, such as KYT forms, perhaps in higher dimensions.  Since the Kaluza--Klein lift on the sphere $S^5$ is much more complicated, leading to a 10-dimensional solution whose only matter field is the self-dual 5-form field strength, there may be more intricate geometric structures involved.


\section*{Acknowledgements}


This work was supported in part by the European Union's Seventh Framework Programme under grant agreement (FP7-REGPOT-2012-2013-1) no.\ 316165.

\appendix


\section{Metric inverse of 3-charge Wu solution}


The 7-parameter charged, rotating black hole solution of gauged supergravity \cite{Wu:2011gq} has metric inverse components $g^{a b} = H_3^{-1} \rho^{-2} \overline{g}^{a b}$ given by
\begin{align}
\overline{g}^{t t} & = - \fr{\Xi_a \Xi_b}{g^2 \Delta_\theta} + \fr{\Xi_a \Xi_b (r^2 + a^2) (r^2 + b^2)}{g^2 \Delta_r} - \fr{8 m^2}{\Delta_r} a b s_{1 2 3} c_{1 2 3} c_{1 2 3 a} c_{1 2 3 b} \nnr
& \quad - \fr{2 m}{g^2 \Delta_r} c_{1 2 3}^2 [r^2 + a^2 b^2 g^2 - 2 m a^2 b^2 g^4 (s_{1 2}^2 + s_{2 3}^2 + s_{3 1}^2) - 4 m a^4 b^4 g^8 s_{1 2 3}^2 ] \nnr
& \quad + \fr{2 m}{g^2 \Delta_r} \Xi_a \Xi_b \{ r^2 [ c_{1 2 3}^2 - 1 + (1 - \Xi_a \Xi_b) s_{1 2 3}^2 ] \nnr
& \quad - a^2 b^2 g^2 [s_{1 2}^2 + s_{2 3}^2 + s_{3 1}^2 + (1 - a^2/b^2 - b^2/a^2 + a^2 g^2 + b^2 g^2) s_{1 2 3}^2] \} \nnr
& \quad - \fr{4 m^2}{g^2 \Delta_r} s_{1 2 3}^2 \Xi_a \Xi_b [1 - \Xi_a \Xi_b + a^2 b^2 g^4 (c_{1 2}^2 + c_{2 3}^2 + c_{3 1}^2)] ,
\end{align}
\begin{align}
\overline{g}^{t \phi_1} & = - \fr{2 m}{\Delta_r} a c_{1 2 3} c_{1 2 3 a} [r^2 + b^2 - 2 m b^2 g^2 (s_{1 2}^2 + s_{2 3}^2 + s_{3 1}^2) - 4 m b^4 g^4 s_{1 2 3}^2] \nnr
& \quad + \fr{2 m}{\Delta_r} b s_{1 2 3} c_{1 2 3 b} \{ \Xi_a^2 [r^2 + b^2 + 2 m (1 + 1/a^2 g^2 + s_1^2 + s_2^2 + s_3^2)] \nnr
& \quad - 2 m [a^4 g^4 c_{1 2 3}^2 + (1/a^2 g^2) c_{1 2 3 a}^2] \} ,
\end{align}
\begin{align}
\overline{g}^{t \phi_2} & = - \fr{2 m}{\Delta_r} b c_{1 2 3} c_{1 2 3 b} [r^2 + a^2 - 2 m a^2 g^2 (s_{1 2}^2 + s_{2 3}^2 + s_{3 1}^2) - 4 m a^4 g^4 s_{1 2 3}^2] \nnr
& \quad + \fr{2 m}{\Delta_r} a s_{1 2 3} c_{1 2 3 a} \{ \Xi_b^2 [r^2 + a^2 + 2 m (1 + 1/b^2 g^2 + s_1^2 + s_2^2 + s_3^2)] \nnr
& \quad - 2 m [b^4 g^4 c_{1 2 3}^2 + (1/b^2 g^2) c_{1 2 3 b}^2] \} ,
\end{align}
\begin{align}
\overline{g}^{\phi_1 \phi_1} & = \Xi_a \sin^2 \theta - \fr{\Xi_a (a^2 - b^2) (1 + g^2 r^2) (r^2 + b^2)}{\Delta_r} - \fr{8 m^2}{\Delta_r} a b g^2 s_{1 2 3} c_{1 2 3} c_{1 2 3 a} c_{1 2 3 b} \nnr
& \quad - \fr{2 m}{\Delta_r} c_{1 2 3 a}^2 [ b^2 + a^2 g^2 r^2 - 2 m b^2 g^2 (s_{1 2}^2 + s_{2 3}^2 + s_{3 1}^2) - 4 m (b^4 g^2 / a^2) s_{1 2 3}^2 ] \nnr
& \quad - \fr{2 m}{a^2 \Delta_r} \Xi_a (a^2 - b^2) \{ r^2 [ c_{1 2 3 a}^2 - 1 + a^2 b^2 g^4 (a^2 / b^2 + a^2 g^2 - 1) s_{1 2 3}^2 ] \nnr
& \quad - a^2 b^2 g^2 [s_{1 2}^2 + s_{2 3}^2 + s_{3 1}^2 + (1 - a^2/b^2 + a^2 g^2 + b^2 g^2 - a^2 b^2 g^4) s_{1 2 3}^2] \} \nnr
& \quad + \fr{4 m^2 b^2 g^2}{a^2 \Delta_r} \Xi_a (a^2 - b^2) s_{1 2 3}^2 (a^2/b^2 + a^2 g^2 - 1 + c_{1 2 a}^2 + c_{2 3 a}^2 + c_{3 1 a}^2) ,
\end{align}
\begin{align}
\overline{g}^{\phi_1 \phi_2} & = - \fr{2 m}{\Delta_r} a b c_{1 2 3 a} c_{1 2 3 b} [1 + g^2 r^2 - 2 m g^2 (s_{1 2}^2 + s_{2 3}^2 + s_{3 1}^2) - 4 m g^2 s_{1 2 3}^2] \nnr
& \quad + \fr{2 m}{\Delta_r} s_{1 2 3} c_{1 2 3} \{ (a^2 - b^2)^2 g^4 [r^2 + 1/g^2 + 2 m (1/a^2 g^2 + 1/b^2 g^2 + s_1^2 + s_2^2 + s_3^2)] \nnr
& \quad - 2 m g^2 [(b^4/a^2) c_{1 2 3 a}^2 + (a^4/b^2) c_{1 2 3 b}^2] \} ,
\end{align}
\begin{align}
\overline{g}^{\phi_2 \phi_2} & = \Xi_b \cos^2 \theta - \fr{\Xi_b (b^2 - a^2) (1 + g^2 r^2) (r^2 + a^2)}{\Delta_r} - \fr{8 m^2}{\Delta_r} a b g^2 s_{1 2 3} c_{1 2 3} c_{1 2 3 a} c_{1 2 3 b} \nnr
& \quad - \fr{2 m}{\Delta_r} c_{1 2 3 b}^2 [ a^2 + b^2 g^2 r^2 - 2 m a^2 g^2 (s_{1 2}^2 + s_{2 3}^2 + s_{3 1}^2) - 4 m (a^4 g^2 / b^2) s_{1 2 3}^2 ] \nnr
& \quad - \fr{2 m}{b^2 \Delta_r} \Xi_b (b^2 - a^2) \{ r^2 [ c_{1 2 3 b}^2 - 1 + a^2 b^2 g^4 (b^2 / a^2 + b^2 g^2 - 1) s_{1 2 3}^2 ] \nnr
& \quad - a^2 b^2 g^2 [s_{1 2}^2 + s_{2 3}^2 + s_{3 1}^2 + (1 - b^2/a^2 + a^2 g^2 + b^2 g^2 - a^2 b^2 g^4) s_{1 2 3}^2] \} \nnr
& \quad + \fr{4 m^2 a^2 g^2}{b^2 \Delta_r} \Xi_b (b^2 - a^2) s_{1 2 3}^2 (b^2/a^2 + b^2 g^2 - 1 + c_{1 2 b}^2 + c_{2 3 b}^2 + c_{3 1 b}^2) ,
\end{align}
\begin{align}
\overline{g}^{r r} & = \Delta_r / r^2 , & \overline{g}^{\theta \theta} & = \Delta_\theta .
\end{align}
We have denoted the constants
\begin{align}
\Xi_a & = 1 - a^2 g^2 , & \Xi_b & = 1 - b^2 g^2 , & c_{I a} & = \sr{1 + a^2 g^2 s_I^2} , & c_{I b} & = \sr{1 + b^2 g^2 s_I^2} , \nnr
s_{I J} & = s_I s_J , & c_{I J} & = c_I c_J , & s_{1 2 3} & = s_1 s_2 s_3 , & c_{1 2 3} & = c_1 c_2 c_3 , \nnr
c_{I J a} & = c_{I a} c_{J a} , & c_{I J b} & = c_{I b} c_{J b} , & c_{1 2 3 a} & = c_{1 a} c_{2 a} c_{3 a} , & c_{1 2 3 b} & = c_{1 b} c_{2 b} c_{3 b} ,
\label{constants}
\end{align}
and the latitudinal function
\be
\Delta_\theta = 1 - g^2 (a^2 \cos^2 \theta + b^2 \sin^2 \theta) ,
\label{Deltatheta}
\ee
while $\Delta_r$ is a complicated function of $r$ that we do not repeat here.


\end{document}